\documentclass[aps,prd,preprint,superscriptaddress,tightenlines,nofootinbib]{revtex4}

\usepackage{graphicx}% Include figure files
\usepackage{dcolumn}% Align table columns on decimal point
\usepackage{bm}% bold math

\begin{document}

\preprint{CLEO CONF 06-12}   % For conference papers

\title{$D^0\bar D^0$ Quantum Correlations, Mixing, and Strong Phases}
\thanks{Submitted to the 33$^{\rm rd}$ International Conference on High Energy
Physics, July 26 - August 2, 2006, Moscow}

\author{D.~M.~Asner}
\author{K.~W.~Edwards}
\affiliation{Carleton University, Ottawa, Ontario, Canada K1S 5B6}
\author{R.~A.~Briere}
\author{I.~Brock~\altaffiliation{Current address: Universit\"at Bonn; Nussallee 12; D-53115 Bonn}}
\author{J.~Chen}
\author{T.~Ferguson}
\author{G.~Tatishvili}
\author{H.~Vogel}
\author{M.~E.~Watkins}
\affiliation{Carnegie Mellon University, Pittsburgh, Pennsylvania 15213}
\author{J.~L.~Rosner}
\affiliation{Enrico Fermi Institute, University of
Chicago, Chicago, Illinois 60637}
% ...
%\author{(CLEO Collaboration)} %FOR PRD_SPECIAL_CHANGEME
\collaboration{CLEO Collaboration} %FOR PRL,CLNS
\noaffiliation
%please hard code the date when you have a final draft and submit to CLEOAC
\date{July 24, 2006}

\begin{abstract} 
% Insert abstract here.
Due to the quantum correlation between the pair-produced $D^0$ and
$\bar D^0$ from the decay of the
$\psi(3770)$, the time-integrated single and double tag decay rates depend
on charm mixing amplitudes, doubly-Cabibbo-suppressed amplitudes, and the
relative strong phase $\delta$ between $D^0$ and $\bar D^0$ decays to
identical final states.  Using 281 ${\rm pb}^{-1}$ collected with the
CLEO-c detector on the $\psi(3770)$ resonance, we measure the absolute
branching fractions of $D^0$ decays to hadronic flavored states, $CP$
eigenstates, and semileptonic final states to determine the relative
strong phase, $\cos\delta$, of the $K^-\pi^+$ final state and to limit the
mixing amplitude $y$.
The results presented in this document are preliminary.
\end{abstract}

\maketitle

When $D^0$ and $\bar D^0$ mesons are pair-produced in $e^+e^-$ collisions
with no accompanying particles (such as through the $\psi(3770)$ resonance),
they are in a
quantum-coherent $C=-1$ state.  Because the initial state (the virtual photon)
has $J^{PC} = 1^{--}$, there follows a set of selection rules for the decays
of the $D^0$ and
$\bar D^0$~\cite{Goldhaber:1976fp,Bigi:1986dp,Bigi:1986rj,Bigi:1989ah,Xing:1996pn,Gronau:2001nr,Atwood:2002ak}.
For example, both $D^0$ and $\bar D^0$
cannot decay to $CP$ eigenstates with the same eigenvalue.  On the other hand,
decay rates to $CP$ eigenstates of opposite eigenvalue are enhanced by a
factor of two.
More generally, final states that can be reached by both $D^0$ and $\bar D^0$
are subject to similar interference effects.  As a result,
the apparent $D^0$ branching fractions in this $D^0\bar D^0$ system differ
from those of isolated $D^0$ mesons.  Moreover, using time-independent
rate measurements, it is possible to probe the $D^0$-$\bar D^0$ mixing
parameters $x\equiv\Delta M/\Gamma$ and $y\equiv\Delta\Gamma/2\Gamma$,
which are the mass and width differences between $D_{CP+}$
and $D_{CP-}$, as
well as the relative strong phases between $D^0$ and $\bar D^0$ decay
amplitudes to any given final state.

We implement the technique presented in Ref.~\cite{Asner:2005wf}, where four
types of final states are considered: flavored
(labeled by $f$ and $\bar f$), $CP+$ eigenstates ($S_+$), $CP-$ eigenstates
($S_-$), and semileptonic ($\ell^+$ and $\ell^-$).  Event yields are
functions of the number of $D^0\bar D^0$ pairs produced (denoted by ${\cal N}$,
branching fractions (denoted by ${\cal B}$), the mixing parameters $y$ and
$R_M\equiv (x^2+y^2)/2$, and the
amplitude ratio $\langle f|\bar D^0\rangle / \langle f|D^0\rangle$,
whose magnitude and phase are denoted by $r_f$ and $-\delta_f$, respectively.
We define $z_f\equiv 2\cos\delta_f$ and give expressions for these yields in
Table~\ref{tab:rateExpressions}, to leading order in $x$ and $y$.

\begin{table}
\begin{tabular}{c|cccc}
\hline
& $f$ & $\ell^+$ & $S_+$ & $S_-$ \\
\hline
$f$        & ${\cal N}{\cal B}_f^2 R_M[1+r_f^2(2-z^2)+r_f^4]$ \\
$\bar f$   & ${\cal N}{\cal B}_f^2 [1+r_f^2(2-z^2)+r_f^4]$ \\
$\ell^-$ & ${\cal N}{\cal B}_f {\cal B}_\ell$ &
        ${\cal N}{\cal B}_\ell^2$ \\
$S_+$      & ${\cal N}{\cal B}_f {\cal B}_{S_+}(1+r_f^2+r_fz_f)$ &
        ${\cal N}{\cal B}_\ell{\cal B}_{S_+}$ & 0 \\
$S_-$      & ${\cal N}{\cal B}_f {\cal B}_{S_-}(1+r_f^2-r_fz_f)$ &
        ${\cal N}{\cal B}_\ell{\cal B}_{S_-}$ &
        $4{\cal N}{\cal B}_{S_+}{\cal B}_{S_-}$ & 0 \\
\hline
$X$ &
        ${\cal N}{\cal B}_f (1+r_f^2 + r_fz_f y)$ &
        ${\cal N}{\cal B}_\ell$ &
        $2{\cal N}{\cal B}_{S_+}(1-y)$ &
        $2{\cal N}{\cal B}_{S_-}(1+y)$ \\
\hline
\end{tabular}
\caption{ST and DT yields for $C=-1$ $D^0\bar D^0$ events, to leading
order in $x$ and $y$.}
\label{tab:rateExpressions}
\end{table}

Our analysis uses 281 ${\rm pb}^{-1}$ of $e^+e^-$ collisions, taken on the
$\psi(3770)$ resonance, with $\sqrt{s} = 3773$ MeV.  The data were
collected with the CLEO-c detector, which is a modification of
CLEO III~\cite{Kubota:1992ww,Hill:1998ea,cleoiiidr,cleorich}, in which the
silicon-strip
vertex detector was replaced with a six-layer vertex drift chamber, whose
wires are all at small stereo angles to the beam axis~\cite{cleocyb}.
The hadronic final states we reconstruct are
$K^-\pi^+$ ($f$), $K^+\pi^-$ ($\bar f$), $K^-K^+$ ($CP+$),
$\pi^+\pi^-$ ($CP+$), $K^0_S\pi^0\pi^0$ ($CP+$), and $K^0_S\pi^0$ ($CP-$).
In the case of the two flavored
final states, $K^-\pi^+$ and $K^+\pi^-$, both of these can be
reached via Cabibbo-favored (CF) or doubly-Cabibbo-suppresssed (DCS)
transitions.  The strong phase between the CF and DCS decay amplitudes,
$\delta_{K\pi}$, is a source of ambiguity in some previous studies of
$D^0$-$\bar D^0$ mixing~\cite{asner}.
We measure yields of both single tags (ST), which are single
fully-reconstructed
$D^0$ or $\bar D^0$ candidates, and double tags (DT), which are events where
both the $D^0$ and $\bar D^0$ are reconstructed.
We identify hadronic $D$ candidates by their beam-constrained mass,
$M \equiv\sqrt{E_{\rm beam}^2 - {\mathbf p}_D^2}$, and by
$\Delta E\equiv E_D - E_{\rm beam}$.  

We also measure semileptonic DT yields, where one $D$ is
fully reconstructed in one of the above hadronic modes
and the other $D$ is required to be semileptonic.  We do not
reconstruct semileptonic single tags because of the undetected
neutrino.  We also omit the DT modes where both $D^0$ and $\bar D^0$
decay semileptonically.
To maximize efficiency, we use inclusive, partial reconstruction of the
semileptonic $D$, demanding that only the electron be found.
When the electron is accompanied by a flavor tag ($K^-\pi^+$ or
$K^+\pi^-$), we further require that the electron and kaon charges be the same,
forming a Cabibbo-favored DT sample.  Doing so
reduces the dominant electron backgrounds, $\gamma\to e^+e^-$ and
$\pi^0\to e^+e^-\gamma$, which are charge-symmetric.  Such a requirement is
unavailable for $CP$-eigenstate tags because they are unflavored.

Efficiencies, backgrounds, and crossfeed among signal modes, are determined
from Monte Carlo (MC) simulations.  
Following the least-squares procedure described in Ref.~\cite{Sun:2005ip},
we perform a fit to these efficiency-corrected yields to extract the
free parameters listed above.  We assume that $K^0_S$ is a purely $CP$-even
eigenstate and that $CP$ violation in $D^0$ decays is negligible.
In Table~\ref{tab:DataResults}, we show the {\it preliminary} results of
the data fit.
Because the precision of the world average for $r_{K\pi}^2$ far exceeds
our determination~\cite{pdg,Abe:2004sn,Link:2004vk}, we constrain this
parameter to be $(3.74\pm 0.18)\times 10^{-3}$ in the fit.
The $\chi^2$ is 15.7 for 20 degrees of freedom, and only
statistical uncertainties have been included.  Systematic uncertainties are
being evaluated, and it is expected that they will be of similar size.
The value of ${\cal B}(D^0\to K^0_S\pi^0)$ shown in
Table~\ref{tab:DataResults} is equivalent to and correlated with the
so-called ``single tag'' measurement in Ref.~\cite{cleo-conf-06-11} of
$(1.212\pm 0.016\pm 0.039)\%$, which is based on the same dataset as
the current analysis but makes use of independently-performed measurements
of $y$~\cite{pdg} and ${\cal N}$~\cite{dhad}, whereas we allow both of
these parameters to be determined by our fit.

As discussed in Ref.~\cite{Asner:2005wf}, systematic effects that are
correlated by final state, such as mismodeling of tracking or $\pi^0$
reconstruction efficiency, cancel in the DCS and mixing parameters.
However, one important source of uncertainty is the quantum-number purity
of the reconstructed $CP$ eigenstates.  Peaking backgrounds to $CP$
eigenstates may come from
flavored decays or $CP$ eigenstates of the opposite eigenvalue.  Therefore,
the size of the simulated background, which assumes
uncorrelated decay, may differ from reality because the quantum correlation
modifies the rates of each of these processes in a different way, and a
systematic uncertainty can be assigned based on the fit results.

Also, the purity of the $C=-1$ initial state may be diluted by
radiated photons, which would reverse the $C$ eigenvalue.  We limit this effect
by searching for DT modes with same-sign $CP$ eigenstates (such as $K^-K^+$ vs.
$\pi^+\pi^-$).  These decays are forbidden for $C=-1$ but are maximally
enhanced for $C=+1$.  Including these yield measurements
(all of which are consistent with zero) and fitting all the other yields to
a sum of $C=-1$ and $C=+1$ contributions, we find no evidence for $C=+1$
contamination --- the $C=+1$ fraction of the sample is $0.06\pm 0.05$ (stat.)
--- and we observe no significant shifts in the fitted parameters.

In summary, using 281 ${\rm pb}^{-1}$ of $e^+e^-$ collisions produced on the
$\psi(3770)$ at CLEO-c, we have searched for $D^0$-$\bar D^0$ mixing and made
a first measurement of the strong phase, $\delta_{K\pi}$.  We expect future
improvements
with the addition of more $CP$ eigenstate modes, more $\psi(3770)$ data, and
higher-energy data with $D^0\bar D^0\gamma$ events, where the
$D^0\bar D^0$ pair is a $C=+1$ eigenstate.

\begin{table}
\begin{tabular}{ccc}
\hline
Parameter & Fitted Value & PDG~\cite{pdg} \\
\hline
${\cal N}$ &
        $(1.09\pm 0.04)\times 10^6$ &
        --- \\
$y$ &
        $-0.058\pm 0.066$ &
        $0.008\pm 0.005$ \\
$R_M$ &
        $(1.7\pm 1.5)\times 10^{-3}$ &
        $< {\cal O}(10^{-3})$ \\
$\cos\delta_{K\pi}$ &
        $1.09\pm 0.66$ &
        --- \\
${\cal B}(D^0\to K^-\pi^+)$ &
        $0.0367\pm 0.0012$ &
        $0.0380\pm 0.0009$ \\
${\cal B}(D^0\to K^-K^+)$ &
        $0.00354\pm 0.00028$ &
        $0.00389\pm 0.00012$ \\
${\cal B}(D^0\to \pi^-\pi^+)$ &
        $0.00125\pm 0.00011$ &
        $0.00138\pm 0.00005$ \\
${\cal B}(D^0\to K^0_S\pi^0\pi^0)$ &
        $0.0095\pm 0.0009$ &
        $0.0089\pm 0.0041$ \\
${\cal B}(D^0\to K^0_S\pi^0$) &
        $0.0127\pm 0.0009$ &
        $0.0155\pm 0.0012$ \\
${\cal B}(D^0\to X e^+\nu_e)$ &
        $0.0639\pm 0.0018$ &
        $0.0687\pm 0.0028$ \\
\hline
\end{tabular}
\caption{Preliminary results from the data fit, with $r_{K\pi}^2$
constrained to be $(3.74\pm 0.18)\times 10^{-3}$.  Uncertainties
on the fit results are statistical only.}
\label{tab:DataResults}
\end{table}

We gratefully acknowledge the effort of the CESR staff
in providing us with excellent luminosity and running conditions.
D.~Cronin-Hennessy and A.~Ryd thank the A.P.~Sloan Foundation.
This work was supported by the National Science Foundation,
the U.S. Department of Energy, and
the Natural Sciences and Engineering Research Council of Canada.

\end{document}